\documentclass[aip,preprint]{revtex4-2}
\usepackage[cp1251]{inputenc}
\usepackage[T2A]{fontenc}
\usepackage[english]{babel}
\usepackage{amssymb,latexsym,amsmath,amscd}
\usepackage{graphicx,color,framed}
\usepackage{xcolor}
\usepackage{siunitx}
\usepackage{empheq}
\usepackage{xr}
\usepackage{microtype}
\usepackage{amsfonts}
\usepackage[normalem]{ulem}
\usepackage{caption}
\usepackage{subcaption}

\begin{document}
\title{Dielectric mismatch effects on polyelectrolyte solutions in electrified nanopores: Insights from mean-field theory}
\author{\firstname{Yury A.} \surname{Budkov}}
\email[]{ybudkov@hse.ru}
\affiliation{Laboratory for Computational Physics, HSE University, Tallinskaya st. 34, 123458 Moscow, Russia}
\affiliation{G.A. Krestov Institute of Solution Chemistry of the Russian Academy of Sciences, 153045, Akademicheskaya st. 1, Ivanovo, Russia}
\author{\firstname{Nikolai N.} \surname{Kalikin}}
\affiliation{Laboratory for Computational Physics, HSE University, Tallinskaya st. 34, 123458 Moscow, Russia}
\affiliation{G.A. Krestov Institute of Solution Chemistry of the Russian Academy of Sciences, 153045, Akademicheskaya st. 1, Ivanovo, Russia}

\begin{abstract}
We utilize self-consistent field theory to explore the mechanical and electrical properties of charged surfaces immersed in polyelectrolyte solutions that could potentially be useful for electrochemical applications. Our research focuses on how the dielectric heterogeneity of the solution could affect the disjoining pressure and differential capacitance of the electric double layer. Relying on the developed theoretical framework, based on the Noether's theorem, we calculate the stress tensor, containing the term, arising from the conformational entropy of the polymer chains. With its help we compute the disjoining pressure in polyelectrolyte solution confined between two parallel charged surfaces and analyze its behavior as a function of separation between the surfaces for different values of dielectric mismatch parameter. We also calculate the differential capacitance of the electric double layer and discuss how dielectric heterogeneity of the polyelectrolyte solution influences its values.
\end{abstract}
\keywords{self-consistent field theory, polyelectrolyte solution, disjoining pressure, differential capacitance}
\maketitle

\section{Introduction}
The description of the polyelectrolyte solution behavior near the charged surfaces is a challenging task due to a complex interplay of electrostatic, steric, and entropic interactions. However, understanding the effects, occurring in such systems may find a broad range of applications in a various fields from material science to biophysics~\cite{dobrynin2005theory,netz2003neutral}.

Modern electrochemical devices, such as batteries and supercapacitors, make extensive use of porous electrodes that are impregnated with low molecular weight electrolyte solutions or room temperature ionic liquids \cite{Nishimura2014,kornyshev2007double,fedorov2014ionic_2}. However, considering long polyelectrolyte chains, one can expect a higher electric double layer (EDL) charge. Recently, we proposed a theoretical model that describes charged polymer chains near an electrified surface~\cite{budkov2022electrochemistry}. Our theoretical findings show a significant increase of differential capacitance values when these chains are in a solution with a polar organic solvent.

From the perspective of possible electrochemical applications~\cite{kalikin2022}, one should emphasize not only the differential capacitance of the EDL but also essential mechanical quantities such as disjoining pressure for discussing slit pores or solvation pressure for pores of arbitrary geometries~\cite{kolesnikov2022electrosorption,kolesnikov2021models}.

The characteristic feature of the polyelectrolyte solution near the charged surfaces is the presence of disjoining pressure oscillations {} with the change of the distance between the walls. Such a behavior was observed experimentally in ref.~\cite{asnacios1997structural}, where the authors studied thin liquid films formed from the semidilute polyelectrolyte solution. They also observed the disappearance of the oscillations with the addition of salt as a result of the additional screening of the electrostatic forces. Similar behavior was studied in ref.~\cite{salmi2007surface} for cationic polyelectrolyte solutions placed between cellulose surfaces. It was shown that the surface forces exhibit oscillating behavior depending on the separation between the surfaces, which is influenced by the molecular weight and charge density of the polymers.

In paper~\cite{yethiraj1999forces} author provides a theoretical investigation of the forces between hard walls immersed in the polyelectrolyte solution within the integral equation framework. The theory predicts oscillatory behavior of the surface force with a period of oscillation depending on the polymer concentration. Studying how the amount of salt added to the solution affects surface forces behavior revealed that it can cause a qualitative shift from predominantly attractive to purely repulsive regimes.

A self-consistent field theory, modeling the depletion of the polyelectrolyte macroions from the area between the charged with the same sign interacting surfaces immersed in the solvent, which also contains electolyte ions, was proposed in ref.~\cite{tadmor2002debye}. The authors explain the ejection of the polymer from a confined space into the bulk reservoir as a result of the competition between the translational and conformational entropies, as well as a consequence of the electrostatic repulsion caused by the polyelectrolyte and surfaces having the same charge sign. The authors proposed an expression for grand potential of the system, which was then used to compute the equilibrium ion concentrations, disjoining pressure between the surfaces and effective Debye length.

In a computer simulation of a system, consisting of two negatively charged interacting surfaces immersed in a solution containing counterions and positively charged polyelectrolytes, the authors of the ref.~\cite{turesson2008simulations} discovered a non-monotonic behavior of the force between the surfaces. The computed osmotic pressure profile demonstrates a steep repulsion at very short separations, followed by the attraction, which the authors attribute to the entropicaly-driven effect of the formation of the polyelectrolyte bridges across the slit from one surface to the other~\cite{aakesson1989electric,podgornik1990forces}.

It is worth noting that in the mentioned studies the disjoining pressure was obtained via the differentiation of the grand potential with respect to the separation between the walls. Recently~\cite{budkov2023macroscopic} we proposed a self-consistent field theory of macroscopic forces in spatially inhomogeneous equilibrium polyelectrolyte solutions and derived the total stress tensor consistent with obtained self-consistent field equations. Such tensor was discovered to include the term, arising from the conformational entropy of flexible polymer chains. Utilizing the expression for this tensor, we calculated the disjoining pressure between two charged surfaces, immersed in the polyelectrolyte solution.

On the other hand, strictly speaking, one can expect that the polyelectrolyte solution has to display a rather significant dielectric heterogeneity due to the difference in the dielectric permittivities of the polymer backbone, solvent and counterions. One can also expect that such phenomenon can strongly influence the properties of the EDL, namely differential capacitance. Thus, in this paper we investigate the effect of the dielectric heterogeneity of the polyelectrolyte solution, considering a dielectric mismatch effect within a simple linear dependence on the concentration, on the disjoining pressure and differential capacitance, basing on the previously developed theoretical background~\cite{budkov2023macroscopic}.

\section{Theoretical background}
\subsection{Self-consistent field equations and stress tensor}
We consider the case of the polyelectrolyte solution consisting of polymerized flexible cations and low-molecular-weight anions, carrying positive $q>0$ and negative $q<0$ charges, correspondingly. The grand thermodynamic potential (GTP) of such system can be written within the developed~\cite{budkov2023macroscopic} notations as follows 
\begin{equation}
\label{GTP}
\Omega=\int d\bold{r}\omega(\bold{r}),
\end{equation}
where we have introduced the GTP density
\begin{equation}
\label{omega_2}
\omega=-\frac{\varepsilon(\nabla\psi)^2}{2}+\rho\psi +f+\frac{k_{B}Tb^2}{6}(\nabla n_{p}^{1/2})^2-\mu_{p}n_{p}-\mu_{c}n_c.
\end{equation}
The first two terms in the expression are the electrostatic energy density in the mean-field approximation with the local charge density $\rho(\bold{r})=q\left(n_{p}(\bold{r})-n_{c}(\bold{r})\right)$
and electrostatic potential $\psi(\bold{r})$, where $n_{p,c}(\bold{r})$ denote the local concentrations of monomeric units and counterions; $\varepsilon$ is the dielectric permittivity of the solution. Considering the difference in the dielectric permittivities of the solvent $\varepsilon_s$, polymerized cations $\varepsilon_p$ and monomeric anions $\varepsilon_c$, we introduce the dielectric mismatch parameters $\delta_p=(\varepsilon_p-\varepsilon_s)/\varepsilon_s$ and $\delta_c=(\varepsilon_c-\varepsilon_s)/\varepsilon_s$. Thus, the introduced above dielectric permittivity of the solution can be written as~\cite{khokhlov1994polyelectrolyte,khokhlov1996weakly,kramarenko2000three,kramarenko2002influence,budkov2017polymer}
\begin{equation}
\label{diel_mismatch}
\varepsilon=\varepsilon_s\left(1+\delta_p\phi_p+\delta_c\phi_c\right),    
\end{equation}
where $\phi_{p,c}=n_{p,c}v$ are the local volume fractions of the monomeric units and counterions. The third term, $f=f(n_{p},n_{c})$, determines the contribution of the volume interactions of monomeric units and counterions to the total free energy density, which is described within the lattice model ~\cite{borukhov1997steric,kornyshev2007double,Maggs2016} as
$f=k_{B}Tv^{-1}\left(\phi_c\ln\phi_{c}+\left(1-\phi_{c}-\phi_{p}\right)\ln\left(1-\phi_{c}-\phi_{p}\right)\right)$; $v$ is the elementary cell volume; note that here we consider only the case of very long polyelectrolyte chains, thereby, neglecting the contribution of their translation entropy to the free energy density. The fourth term is the density of the conformational free energy \cite{lifshitz1969some,khokhlov1994statistical,borue1990statistical} of the flexible polymer chains with a bond length $b$, which determines the volume of the elementary cell in a standard way as $v=b^3$; $k_{B}$ is the Boltzmann constant, $T$ is the temperature. The last terms in the GTP density expression contain the bulk chemical potentials, $\mu_{p,c}$, of the monomeric units and counterions, respectively.

The self-consistent field equations can be obtained as the Euler-Lagrange equations for functional (\ref{GTP})
\begin{equation}
\label{EL_eq}
\frac{\partial{\omega}}{\partial{\psi}}=\partial_{i}\frac{\partial \omega}{\partial(\partial_{i}\psi)},~\frac{\partial{\omega}}{\partial{n_{p}^{1/2}}}=\partial_{i}\frac{\partial \omega}{\partial(\partial_{i}n_{p}^{1/2})},~\frac{\partial{\omega}}{\partial{n_{c}}}=0,
\end{equation}
where $\partial_i=\partial/\partial{x}_{i}$ is the partial derivative with respect to the Cartesian coordinates $x_i$ ($i=1,2,3$) and the standard Einstein summation rule is implied. Using GTP density (\ref{omega_2}) and expression for the dielectric permittivity (\ref{diel_mismatch}), we obtain
\begin{empheq}[left=\empheqlbrace]{align}\nonumber
\label{scf_eq}
&\bar{\mu}_{p}(\bold{r})-\frac{k_{B}Tb^2}{6n_{p}^{1/2}(\bold{r})}\nabla^2 n_{p}^{1/2}(\bold{r})+q\psi(\bold{r})-\frac{\alpha_p(\nabla\psi(\bold{r}))^2}{2}=\mu_{p}\\
&\bar{\mu}_{c}(\bold{r})-q\psi(\bold{r})-\frac{\alpha_c(\nabla\psi(\bold{r}))^2}{2}=\mu_{c}\\\nonumber
&\nabla(\varepsilon\nabla\psi(\bold{r}))=-q\left(n_{p}(\bold{r})-n_{c}(\bold{r})\right),
\end{empheq}
where 
$\alpha_{p,c}=v\delta_{p,c}$ are the effective polarizabilities of the species, and $\bar{\mu}_{c}={\partial f}/{\partial n_{c}}=k_{B}T\ln\left({\phi_c}/{(1-\phi_c-\phi_p})\right)$,
$\bar{\mu}_{p}={\partial f}/{\partial n_{p}}=-k_{B}T\ln\left(1-\phi_{c}-\phi_{p}\right)$
are the intrinsic chemical potentials of monomeric units and counterions, respectively. The first and second equations of the system (\ref{scf_eq}) are the chemical equilibrium conditions~\cite{landau2013statistical} for the monomeric units and counterions, respectively. In contrast to our previous work~\cite{budkov2023macroscopic}, the left hand side of each chemical equilibrium conditions contains additional term $-\alpha_{p,c}(\nabla\psi)^2/2$ that is nothing but a potential energy of induced dipole $\bold{p}_{p,c}=\alpha_{p,c}\boldsymbol{\mathcal{E}}$ in "external" field $\boldsymbol{\mathcal{E}}=-\nabla\psi(\bold{r})$~\cite{hatlo2012electric}. The third equation is the standard Poisson equation for the local electrostatic potential taking into account the dielectric heterogeneity of the solution~\cite{landau2013electrodynamics}. Taking into account that in the bulk solution, where $\psi=0$, the local electroneutrality condition, $n_{p}=n_c=n_0$, is fulfilled, we obtain the following expressions for the bulk chemical potentials of the species 
$\mu_c=k_{B}T\ln\left({\phi_0}/(1-2\phi_0)\right)$, $\mu_p=-k_{B}T\ln\left(1-2\phi_0\right)$,
where $\phi_0=n_0v$ is the bulk volume fraction of the monomeric units and counterions. The boundary conditions for the polymer concentration and electrostatic potential are~\cite{netz2003neutral,landau2013electrodynamics} $n_{p}|_s=0$, $\psi|_{s}=\psi_0$, where the symbol $|_{s}$ means that the variables are calculated at the surfaces of immersed macroscopic conductors. These boundary conditions mean that near the surface of a conductive wall (with a fixed surface potential, $\psi_0$) the monomeric units are exposed to a strong repulsive force~\cite{netz2003neutral}. Note that for simplicity we neglect the specific adsorption of the counterions. The latter can be directly taken into account~\cite{budkov2018theory}.

Referring to our recently developed theoretical framework, based on the application of Noether's theorem-based approach \cite{budkov2022modified,budkov2023macroscopic,brandyshev2023noether}, we can write the stress tensor as follows
\begin{equation}
\sigma_{ik}=\omega\delta_{ik}-\partial_{i}n_{p}^{1/2}\frac{\partial \omega}{\partial(\partial_{k}n_{p}^{1/2})}-\partial_{i}\psi\frac{\partial \omega}{\partial(\partial_{k}\psi)},
\end{equation}
which after some algebra can be split into three contributions
\begin{equation}
\label{stress}
\sigma_{ik}=\sigma^{(h)}_{ik}+\sigma^{(M)}_{ik}+\sigma_{ik}^{(c)},
\end{equation}
where
\begin{equation}
\sigma_{ik}^{(h)}=-P\delta_{ik}
\end{equation}
is the standard hydrostatic stress tensor with the local osmotic pressure $P=n_p\bar{\mu}_p+n_c\bar{\mu}_c-f=-{k_{B}T}v^{-1}\left(\ln\left(1-\phi_p-\phi_c\right)+\phi_p\right),$
\begin{equation}
\sigma_{ik}^{(M)}=\varepsilon\mathcal{E}_{i}\mathcal{E}_{k}-\frac{\varepsilon_s}{2}\mathcal{E}^2\delta_{ik}
\end{equation}
is the electrostatic contribution described by the standard Maxwell stress tensor~\cite{landau2013electrodynamics} for the case of continuous dielectric medium with the electrostatic field components $\mathcal{E}_i=-\partial_{i}\psi$, and
\begin{equation}
\label{conf_stress}
\sigma_{ik}^{(c)}=\frac{k_{B}Tb^2}{3}\left(\frac{1}{2}\nabla\cdot\left(n_p^{1/2}\nabla n_p^{1/2}\right)\delta_{ik}-\partial_{i}n_{p}^{1/2}\partial_{k}n_{p}^{1/2}\right)
\end{equation}
is the contribution to the total stress tensor stemming from the conformational entropy of the polymer chains (so-called {\sl conformational} stress tensor), which was first recently obtained in ref.~\cite{budkov2023macroscopic}.

We would like to note that discovered conformational stress tensor (\ref{conf_stress}) can be used for assessment of surface forces and deformations in various polymeric systems with flexible chains, including polymer solutions, melts, gels, membranes, {\sl etc.}

\subsection{Slit pore case. Disjoining pressure}

For the case of the polyelectrolyte solution confined in a slit nano-sized pore with identical charged conducting walls, placed at $z=0$ and $z=H$ with $H$ being the pore size, and considering the equilibrium between the solution phases in the confinement and in the bulk, we can reduce the self-consistent field equations to the following form
\begin{empheq}[left=\empheqlbrace]{align}\nonumber
\label{scf_eq_in_pore}
&\bar{\mu}_{p}(z)-\frac{k_{B}Tb^2}{6\phi_{p}^{1/2}(z)}\frac{d^2\phi_{p}^{1/2}(z)}{dz^2}+q\psi(z)-\frac{\alpha_p}{2}\left(\frac{d\psi(z)}{dz}\right)^2=\mu_{p}\\
&\frac{d}{dz}\left(\varepsilon\frac{d\psi(z)}{dz}\right)=-\frac{q}{v}\left(\phi_{p}(z)-\phi_{c}(z)\right)\\\nonumber
\end{empheq}
with the boundary conditions $\psi(0)=\psi(H)=\psi_0$,~$\phi_{p}(0)=\phi_{p}(H)=0$; we can also analytically express the volume fraction of the counterions as 
\begin{equation}
    \phi_c=\frac{(1-\phi_p)e^{\frac{\mu_c+q\psi+\frac{\alpha_c\psi'^2}{2}}{k_BT}}}{1+e^{\frac{\mu_c+q\psi+\frac{\alpha_c\psi'^2}{2}}{k_BT}}},    
\end{equation}
{}{where $'$ denotes the derivative with respect to z coordinate.} 
In what follows, we assume that the counterion-induced dielectric mismatch parameter is much smaller than that associated with polymerized cations, which is the case for the solutions of polymeric ionic liquids, i.e. $\alpha_c = 0$.

Using the expression for the stress tensor (\ref{stress}), we can determine the disjoining pressure in the pore. In the case of the slit pore, the mechanical equilibrium condition can be written as
\begin{equation}
\frac{d\sigma_{zz}}{dz}=0,
\end{equation}
thereby, leading to
\begin{equation}
-\sigma_{zz}(z)=P_{b}+\Pi=const,
\end{equation}
where $\sigma_{zz}$ is the normal stress, $P_b$ is the bulk pressure and $\Pi$ is the disjoining pressure~\cite{derjaguin1987derjaguin}. Thus, determining $\sigma_{zz}$ at the middle of the pore at $z=H/2$, where $n_{p}^{\prime}(H/2)=\mathcal{E}(H/2)=0$, we obtain 
$\Pi=-\sigma_{zz}\left({H}/{2}\right)-P_b=-{k_{B}Tb^2}n_p^{\prime\prime}\left({H}/{2}\right)/{12}+P_m-P_b$, where we have introduced the pressure at the pore middle, $P_m=P\left({H}/{2}\right)$. The latter equation can be rewritten in a form that is more useful for applications without the second derivative of the polymer concentration~\cite{budkov2023macroscopic}. Using eq. (\ref{scf_eq_in_pore}) at $z=H/2$, i.e. $-{k_BTb^2}n_p^{\prime\prime}\left({H}/{2}\right)/{12}=n_{pm}\left(\mu_{p}-\bar\mu_{pm}-q\psi_m\right)$, we arrive at
\begin{equation}
\label{disj_press}
\Pi=P_m-P_b+n_{pm}\left(\mu_{p}-\bar\mu_{pm}-q\psi_m\right),
\end{equation}
where $\psi_m=\psi\left({H}/{2}\right)$,~$n_{pm}=n_p\left({H}/{2}\right)$,~$\bar\mu_{pm}=\bar{\mu}_p\left({H}/{2}\right)$.
 
\section{Numerical results}
Now, we turn to the numerical results for the disjoining pressure and differential capacitance. The main goal of the study is to investigate the influence of the dielectric mismatch effect on the profiles of the disjoining pressure and differential capacitance of the EDL. First, we calculate the disjoining pressure. For this we need to solve self-consistent field equations (\ref{scf_eq_in_pore}), calculate the respective variables at the midpoint of the pore and then utilize the eq. (\ref{disj_press}) for different pore widths. Then, we examine the influence of the dielectric mismatch effect on the behavior of differential capacitance, $C=d\sigma/d\psi_0$, as a function of the surface potential, where $\sigma=-\varepsilon \psi^{\prime}(0)$ is the surface charge density of the pore walls~\cite{kornyshev2007double,budkov2021electric}. The following results were computed for temperature $T=300$ K, elementary charge $q=1.6\times10^{-19}$ C, segment length $b=0.5$ nm, bulk volume fraction $\phi_0=0.1$, which represents the case of the polyelectrolyte solution, and the solvent dielectric permittivity $\varepsilon_s=40\varepsilon_0$, which models polar solvents as dimethyl sulfoxide or acetonitrile.

\subsection{Disjoining pressure}
\begin{figure}
     \centering
     \begin{subfigure}[b]{\textwidth}
         \centering
         \includegraphics[width=11 cm]{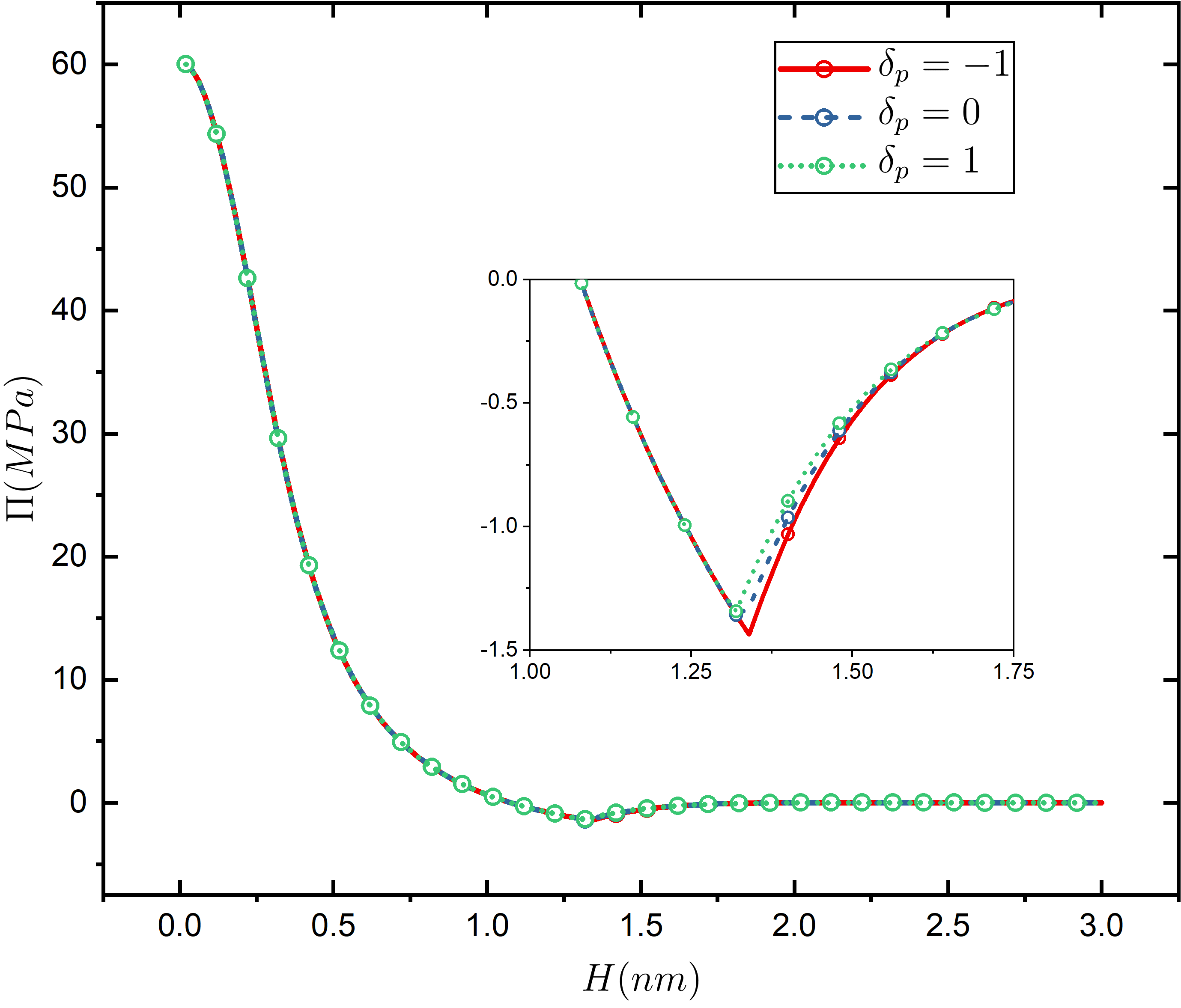}
         \subcaption{}
         \label{fig1:dpress_high_positive}
     \end{subfigure}
     \vfill
     \begin{subfigure}[b]{\textwidth}
         \centering
         \includegraphics[width=11 cm]{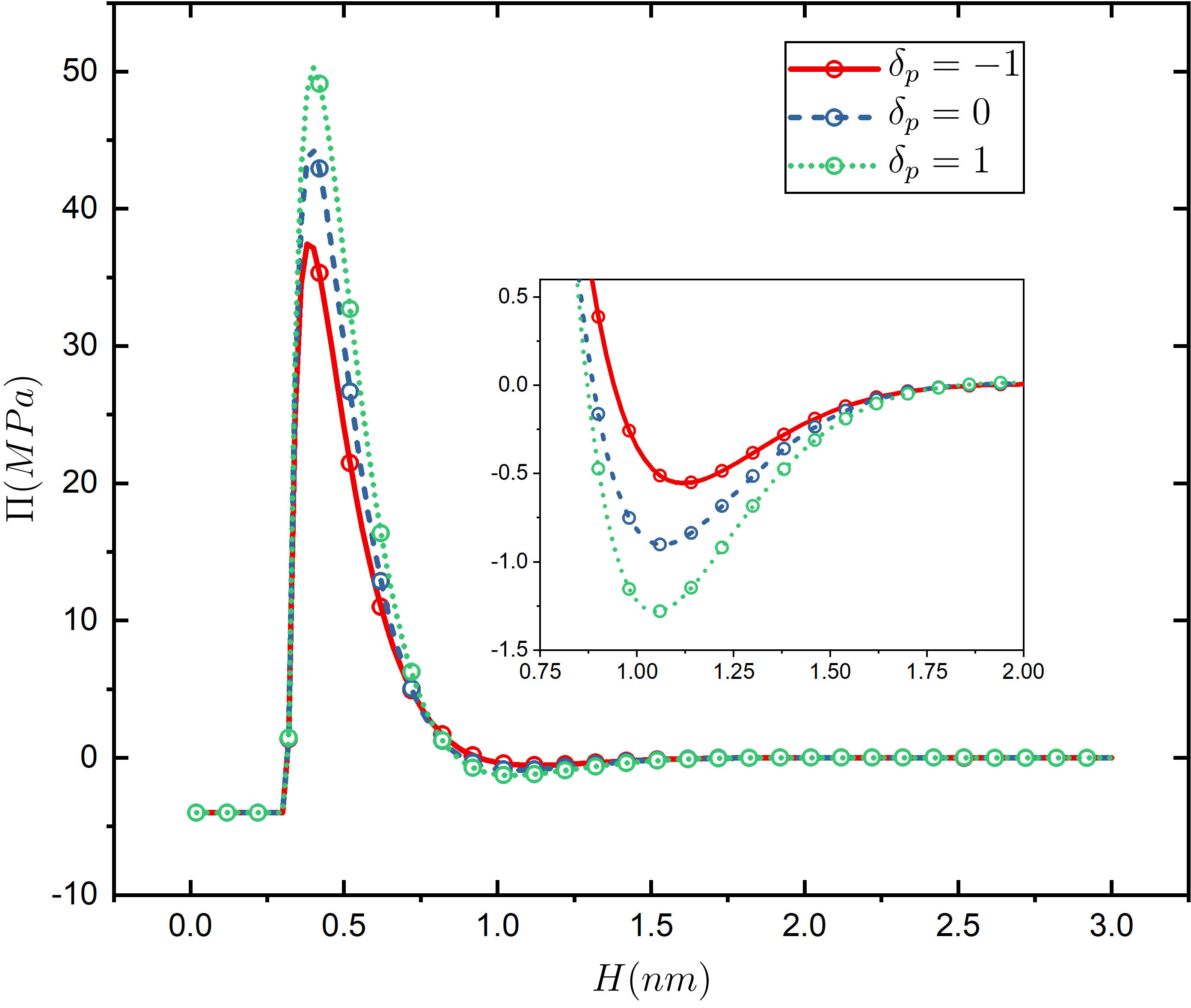}
         \subcaption{}
         \label{fig1:dpress_high_negative}
     \end{subfigure}
     \caption{Disjoining pressure as a function of the distance between the walls for a set of dielectric mismatch parameter $\delta_p$ with positive $\psi_0=0.1~V$ (\ref{fig1:dpress_high_positive}) and negative $\psi_0=-0.1~V$ (\ref{fig1:dpress_high_negative}) surface potentials. The data are shown for $\phi_0=0.1$, $\varepsilon_s=40\varepsilon_0$, $b=v^{1/3}=0.5~nm$, $T=300~K$ and $q=1.6\times 10^{-19}~C$.}
\end{figure}

Fig. \ref{fig1:dpress_high_positive} shows the disjoining pressure as a function of the surfaces separation of the polyelectrolyte solution for rather high positive potential $\psi=0.1~V$, with different values of the dimensionless dielectric mismatch parameter $\delta_p=\alpha_p/v$. As is seen, the disjoining pressure behaves non-monotonically and the position of the minimum determines the width at which the polymer chains start to permeate the pore. Moreover, we can see that disjoining pressure is not very sensitive to the change in the mismatch parameter, leading to a practically unnoticeable shift of the minimum.

The same nonmonotonic trend of the disjoining pressure on the pore width for the case of the negative surface potential is shown in fig.~\ref{fig1:dpress_high_negative}. As is seen, for the case of rather high values of negative surface potential, we observe that as the $\delta_p$ increases, the minimum of the disjoining pressure deepens. Such a behavior agrees with the results of the aforementioned Monte Carlo simulation~\cite{turesson2008simulations}, where the oppositely charged polyelectrolytes and surfaces stimulate the formation of the polymeric bridging~\cite{podgornik1993stretching,podgornik1995colloidal,podgornik2006polyelectrolyte}, displaying the attraction between the surfaces. One may conclude that the change of $\delta_p$ from negative to positive values provokes an increase in polymer concentration in the pores, which explains the deepening of the minimum on the disjoining pressure curve, i.e. the growth of the attractive bridging-induced interaction. It also should be noted that at such values of the surface potential the model predicts that the polymerized cations enter the pore at the width $H\approx 0.25$~nm, which is smaller than assumed bond length $b$. Thus, we can consider that physically meaningful values of the disjoining pressure start approximately from the distance, corresponding to the location of the curve maximum. Considering this remark the obtained disjoining pressure behavior is in agreement with the computer simulations~\cite{turesson2008simulations}.

\subsection{Differential capacitance}
\begin{figure}
     \centering
     \begin{subfigure}[b]{\textwidth}
         \centering
         \includegraphics[width=11 cm]{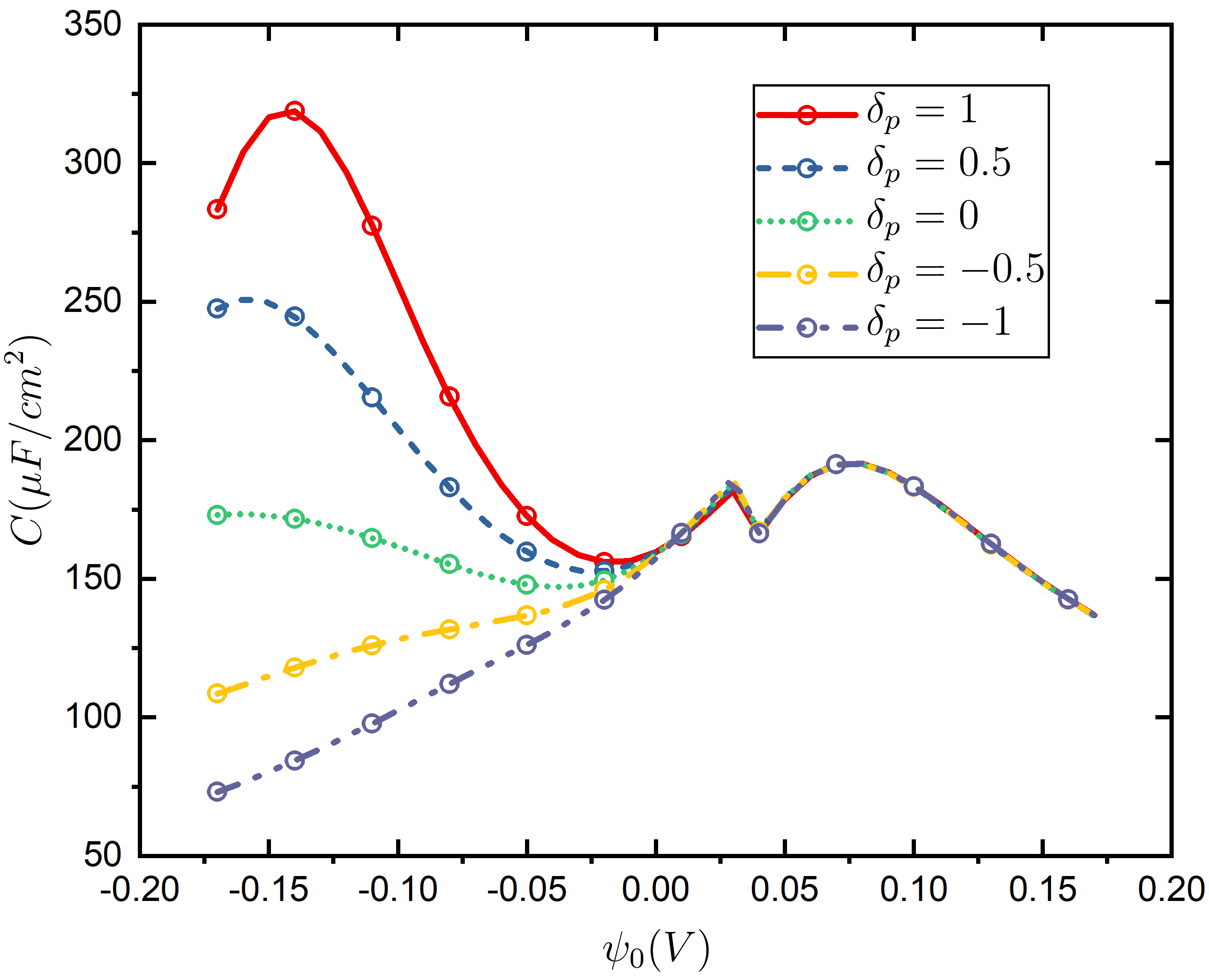}
         \subcaption{}
         \label{fig2:capacitance_1nm}
     \end{subfigure}
     \vfill
     \begin{subfigure}[b]{\textwidth}
         \centering
         \includegraphics[width=11 cm]{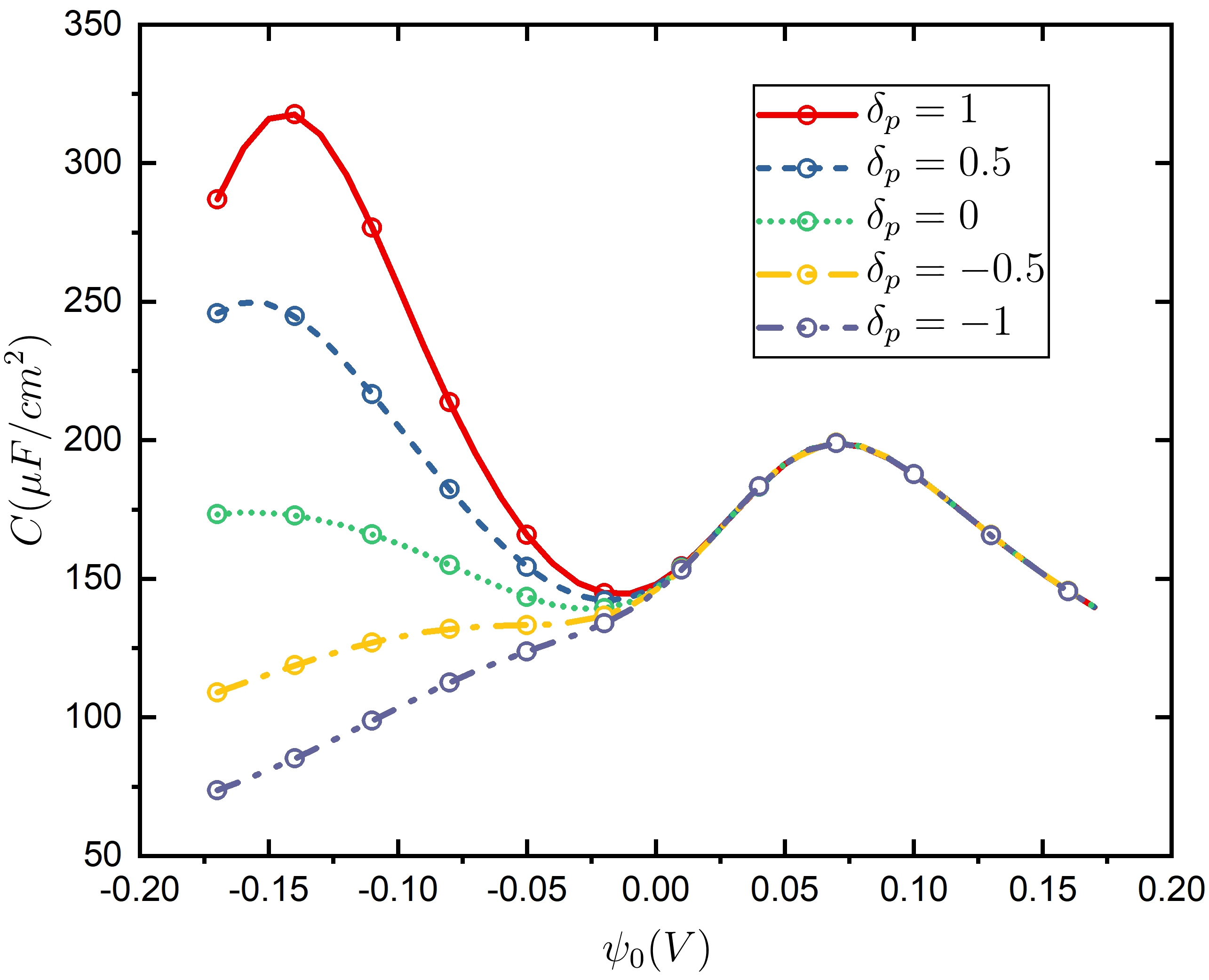}
         \subcaption{}
         \label{fig2:capacitance_2nm}
     \end{subfigure}
     \caption{Differential capacitance profiles for a pore of 1 nm (\ref{fig2:capacitance_1nm}) and 2 nm (\ref{fig2:capacitance_2nm}) width, plotted for different values of the mismatch parameter $\delta_p$. The data are shown for $\phi_0=0.1$, $\varepsilon_s=40\varepsilon_0$, $b=v^{1/3}=0.5~nm$, $T=300~K$ and $q=1.6\times 10^{-19}~C$.}
\end{figure}

Now let us turn to the results of the electric differential capacitance profiles calculation. As one can see at fig. \ref{fig2:capacitance_1nm} and \ref{fig2:capacitance_2nm} the values of the differential capacitance are highly sensitive to the change in the parameter $\delta_p$. The profiles differ only in the region of negative surface potentials due to the fact that at the region of positive potentials polymerized cations are practically absent in the pore due to the electrostatic repulsion. At the increase in the negative surface potentials in the dielectric mismatch parameter, $\delta_p$, results in an increase in the differential capacitance (growth of the maximum differential capacitance). Such behavior can be attributed to the fact that a larger values of $\delta_p$ correspond to a stronger dielectrophoretic force\cite{budkov2015modified2,budkov2016theory2,budkov2018theory,budkov2020two2,budkovJPCC2021_2}, which in turn leads to a more effective "sucking" of the polarizable macromolecules from the bulk solution into the charged pore. It should be noted that this differential capacitance behavior resembles that discovered recently within the theoretical framework for monomeric ionic liquids in ref.~\cite{budkovJPCC2021_2}. 

Moreover, similar to our previous study \cite{budkov2023macroscopic}, in the case of rather small pore of 1 nm width (see fig. \ref{fig2:capacitance_1nm}) we obtain an abrupt jump in differential capacitance values when turning to the region of positive surface potentials. Such behavior is based on the same phenomenon as the ejection of the polymerized cations from the charged slit pore. When the distance between the surfaces increases, the differential capacitance profiles become smooth and resemble those obtained for the case of isolated EDL, as demonstrated for the case of pore of 2 nm width in Fig. \ref{fig2:capacitance_2nm}.

\section{Concluding remarks}
To summarize, we used the developed self-consistent field theory to examine how the mechanical and electrical properties of charged surfaces immersed in a polyelectrolyte solution are affected by the dielectric heterogeneity of the solution. Namely, we observed that the change of the dielectric mismatch parameter $\delta_p$ from negative to positive values for the case of polymerized cations and pore walls with opposite charges leads to a more pronounced minimum on the disjoining pressure curve. Underlying such behavior is the increase of the monomer concentration inside the pore, which leads to the enhancement of the formation of a polyelectrolyte bridge between the walls of the slit. We also obtained a significant increase in differential capacitance values in the region of negative potentials for the same range of values of mismatch parameter $\delta_p$. The explanation here is quite intuitive: As the dielectric mismatch parameter increases, the total dielectric permittivity also increases, leading to reduced screening of the surface potential. This causes an increase in the monomer concentration in the pore, which in turn leads to a higher charge accumulation and an observed growth of the differential capacitance values. 

We believe that our theoretical findings could be relevant for the design of contemporary supercapacitors that make use of polymeric electrolytes, such as polymeric ionic liquids and polyelectrolyte gels, as opposed to low-molecular weight electrolyte solutions and room temperature ionic liquids.

{\bf Acknowledgements.}  
Paper is dedicated to the memory of Igor Ya. Erukhimovich, an outstanding theoretical physicist whose papers had a great influence on the authors. This work is an output of a research project implemented as part of the Basic Research Program at the National Research University Higher School of Economics (HSE University). The numerical calculations were performed on the supercomputer facilities provided by NRU HSE.

\bibliographystyle{aipnum4-2}
\bibliography{lit}

\newpage

\end{document}